\documentclass[lettersize,journal]{IEEEtran}
\usepackage{amsmath,amsfonts}
\usepackage{algorithm}
\usepackage{array}
\usepackage[caption=false,font=normalsize,labelfont=sf,textfont=sf]{subfig}
\usepackage{textcomp}
\usepackage{stfloats}
\usepackage{url}
\usepackage{verbatim}
\usepackage{graphicx}
\usepackage{cite}
\usepackage{algpseudocode}
\usepackage{booktabs}
\usepackage{multirow}
\usepackage[utf8]{inputenc}
\usepackage{amssymb}
\usepackage{makecell}
\usepackage{tabularx}
\usepackage{ragged2e}
\newcolumntype{Y}{>{\RaggedRight\arraybackslash}X}
\usepackage{tikz}
\newcommand*\emptycirc[1][1ex]{\tikz\draw (0,0) circle (#1);} 
\newcommand*\halfcirc[1][1ex]{%
	\begin{tikzpicture}
	\draw[fill] (0,0)-- (90:#1) arc (90:270:#1) -- cycle ;
	\draw (0,0) circle (#1);
	\end{tikzpicture}}
\newcommand*\fullcirc[1][1ex]{\tikz\fill (0,0) circle (#1);}

\begin{document}

\title{Server-Enforced Watermarking in U-Shaped Split Federated Learning}

\author{
    \IEEEauthorblockN{
        Zhengchunmin Dai\IEEEauthorrefmark{1},
        Jiaxiong Tang\IEEEauthorrefmark{1},
        Peng Sun\IEEEauthorrefmark{2},
Honglong~Chen\IEEEauthorrefmark{4},~\IEEEmembership{Senior~Member,~IEEE,} and
        Liantao Wu\IEEEauthorrefmark{1}
    }
\thanks{Zhengchunmin Dai, Jiaxiong Tang and Liantao Wu are with the Shanghai Key Laboratory of Trustworthy Computing (Software Engineering Institute), East China Normal University, Shanghai 200062, China. E-mails: 51275902069@stu.ecnu.edu.cn, 51275902027@stu.ecnu.edu.cn, ltwu@sei.ecnu.edu.cn.}
\thanks{Peng Sun is  with the College of Computer Science and Electronic Engineering, Hunan University, Changsha 410082, China. E-mail: psun@hnu.edu.cn.}
\thanks{Honglong Chen is with the College of Control Science and Engineering, China University of Petroleum (East China), Qingdao 266580, China. Email: chenhl@upc.edu.cn.}
}

\maketitle

\begin{abstract}

U-shaped split federated learning (U-SFL) enables resource-constrained Internet of Things devices to collaboratively train models with an edge server while retaining raw data and labels locally. We consider model-service deployments in which a provider supplies proprietary models whose client-side segments reside on participating devices, creating risks of unauthorized copying and redistribution. Protecting these segments is challenging because the server cannot directly access client data, labels, or model parameters, and potentially malicious clients may refuse to perform watermark embedding. To address this challenge, we propose Sigil, a server-enforced watermarking framework for U-SFL. Sigil defines a secret watermark constraint in the server-visible activation space and embeds the watermark into client-side models by injecting a watermark gradient into the gradients returned during training. This mechanism requires neither access to clients' raw data and labels nor client-side watermarking operations. To limit interference with the main task and reduce detectability by gradient anomaly detectors, Sigil adaptively clips the watermark gradient relative to the main-task gradient. Experiments on two datasets and four model architectures demonstrate high watermark detection rates with limited impact on task accuracy, robustness against the evaluated removal attacks, and stealthiness against the evaluated gradient anomaly detector.

\end{abstract}


\section{Introduction}

Deep Neural Networks (DNNs) underpin a wide range of Internet of Things (IoT) vision services, including visual monitoring and industrial inspection. In these applications, networked cameras and edge gateways collect local data, while edge servers provide computational resources for collaborative model training. The limited computational resources of participating devices and the need to keep sensitive data local motivate training architectures that distribute computation across devices and servers. Meanwhile, developing high-performance vision models requires substantial investment in model design, training, and domain-specific expertise. These models therefore constitute valuable intellectual property (IP) for the organizations that develop and provide them.

However, the high value of these models also makes them targets for attackers, leading to escalating issues of intellectual property rights infringement, such as model leakage, unauthorized copying, and illicit distribution \cite{10251039}. To protect the ownership of DNN models, model watermarking techniques have been proposed and extensively studied. Specifically, existing DNN model watermarking schemes can be broadly categorized into two categories: backdoor watermarking \cite{217591, 10646835} and parameter watermarking \cite{10.1145/3078971.3078974, 287178, 298288}. Backdoor watermarking schemes typically use a secret set of trigger samples to embed ownership information into the target model's input-output behavior, which can later be verified through black-box queries on a suspicious model. In contrast, parameter watermarking schemes encode watermark information, such as a specific bit string, directly into the model's parameter space, which requires white-box access to the model parameters for extraction and verification.

To support collaborative training while keeping raw data local, decentralized machine learning (DCML) paradigms, such as Federated Learning (FL) \cite{MAL-083,laidi2025federated}, Split Learning (SL) \cite{GUPTA20181,2018arXiv181200564V}, and their variants \cite{DBLP:conf/aaai/ThapaCCS22,lin2025adaptsfl}, have received considerable attention. Among these paradigms, SL allows participating devices to offload part of the model computation to a server, making it suitable for collaborative training in resource-constrained IoT deployments \cite{ninkovic2024comsplit}.

Nevertheless, DCML introduces unprecedented challenges for IP protection. On the one hand, involving multiple participants increases the number of potential model-leakage points, thereby heightening the risk of unauthorized use and IP infringement  \cite{DBLP:conf/aaai/LiC0RFC24, ZENG2025107150, 285505}.  On the other hand, differences among participants in computational capabilities, data visibility, model access, and training workflows can render conventional deep neural network (DNN) watermarking methods ineffective or infeasible.

In FL, a central server coordinates global model training, while clients retain their private data and perform model updates locally. Existing FL watermarking methods are generally classified as either client-side \cite{9847383, 9658998, 10.1145/3630636}  or server-side \cite{9603498, 10504977}, depending on their ownership objectives and trust assumptions. Although the server cannot directly access client data, it controls the global model distributed to clients in each training round. Consequently, the server can embed backdoor watermarks using public or noisy data \cite{9603498} or directly encode parameter watermarks into the global model \cite{10504977}. These methods, however, depend on access to the relevant model parameters---an assumption that does not hold for the client-side model segments in the SL setting considered in this work.

SL \cite{GUPTA20181, 2018arXiv181200564V} is another emerging DCML paradigm that partitions a model into client-side and server-side components, allowing collaborative training without transmitting raw client data. U-shaped SL further limits information exposure by retaining both raw data and labels on the client. SL and U-shaped SL can also be integrated with FL to support scalable multi-client training, resulting in Split Federated Learning (SFL) and U-shaped Split Federated Learning (U-SFL), respectively \cite{DBLP:conf/aaai/ThapaCCS22}.

\begin{figure}[t]
  \centering
  \includegraphics[width=0.8\columnwidth]{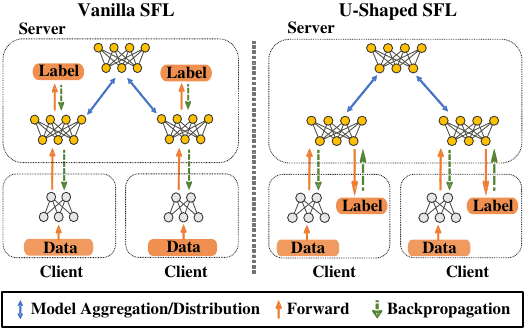}
  \caption{Comparison of Vanilla SFL and U-Shaped SFL Architectures.}
  \label{fig:usfl}
\end{figure}

In the U-SFL workflow shown in Fig.~\ref{fig:usfl}, clients and the server collaboratively train their respective model segments with the server lacking access to the labels. Each client processes its local input data and transmits the resulting intermediate activations, commonly referred to as smashed data, to the server.  The server completes the remaining forward pass and sends the output logits back to the client for loss computation. Gradients are then backpropagated across the split boundary to update both model segments. After several local epochs, the client-side and server-side model parameters are separately aggregated using FedAvg and redistributed for the next round.

\begin{table}[t]
  \centering
  \caption{
    Comparison of server capabilities and watermark embeddability across frameworks.
  }
  \label{tab:have_have_nots}
  \footnotesize
  
  \begin{tabular}{@{}l ccc@{}}
    \toprule
    \textbf{Server Capability} & \textbf{FL} & \textbf{SL/SFL} & \textbf{U-SL/U-SFL} \\
    \midrule
    Client Model Access & \fullcirc & \emptycirc & \emptycirc \\
    Client Labels Access  & \emptycirc & \fullcirc & \emptycirc \\
    Client Data Access         & \emptycirc & \emptycirc & \emptycirc \\
    \midrule
    Parameter Watermark   & \fullcirc & \emptycirc & \emptycirc \\
    Backdoor Watermark  & \halfcirc & \halfcirc & \emptycirc \\
    \bottomrule
  \end{tabular}
  
  \vspace{1ex}
  \begin{tabular}{@{}l@{}}
    \fullcirc\ Full Capability / Embeddable  \\
    \halfcirc\ Partial Capability / Partially Embeddable \\
    \emptycirc\ No Capability / Not Embeddable
  \end{tabular}
\end{table}

We consider a provider-owned edge vision service trained using U-SFL. The model provider operates the server and supplies a proprietary model, while participating IoT devices contribute local training data and update the client-side model segments. U-SFL keeps raw inputs, labels, and loss computation on the devices while offloading part of the model computation to the server. Although the provider retains ownership of the supplied model, participating devices have white-box access to their local model parameters. A malicious participant may therefore copy or redistribute the client-side model segment without authorization. Our objective is to enable the provider to establish verifiable ownership of such client-side model segments.

Protecting client-side model segments is particularly challenging in U-SFL. The server observes intermediate activations and participates in gradient propagation, but it cannot directly access the clients' raw data, labels, or local model parameters. Consequently, the server cannot directly embed a parameter-space watermark into the client-side model or rely on a potentially malicious client to execute the watermarking procedure. This limitation motivates the following research question:

\textit{How can a capability-constrained server embed a verifiable ownership credential into a potentially malicious client's model without accessing the client's raw data, labels, or model parameters?}

To address this challenge, we propose Sigil, a mandatory, server-enforced watermarking framework for capability-constrained servers in U-SFL. Sigil uses the gradient channel to embed the  watermark as a statistical constraint on server-visible activations. Specifically, it encourages the projections of these activations onto a randomly generated secret subspace to match a target bit string. We further introduce an adaptive gradient-clipping mechanism that regulates the watermark gradient to improve embedding fidelity while reducing its detectability by gradient-anomaly detection methods.

Sigil exploits the activation--gradient interface of split learning: the server observes client-generated activations and controls the gradients returned to update the client-side model. This interface allows Sigil to embed the watermark without accessing the clients' raw data, labels, or local model parameters. As summarized in Table~\ref{tab:have_have_nots}, U-shaped SL and U-SFL impose the most restrictive server-side data and model access among the configurations considered. We therefore instantiate and evaluate Sigil in U-SFL to determine whether the activation--gradient interface alone is sufficient for watermark embedding under these capability constraints, while accounting for multi-client training and model aggregation.

We conduct extensive experiments on two representative datasets and four model architectures. The results show that Sigil achieves high watermark fidelity and robustness while maintaining low detectability under the evaluated gradient-anomaly detection method.

In summary, the key innovations and contributions of this research are as follows:
\begin{itemize}
    \item We formalize the distinct IP threat in U-SFL, where the capability-constrained server faces model leakage risks from clients with white-box access to local parameters. To address this threat, we propose Sigil, a server-enforced watermarking framework that embeds the server's ownership credential without access to clients' raw data, labels, or local model parameters during training.
    
    \item We design a watermark-embedding mechanism that utilizes the server-visible activation space and gradient channel to impose a statistical watermark constraint on the client-side model. To reduce detection by potentially malicious clients, we introduce an adaptive gradient clipping mechanism that dynamically regulates the watermark gradient, enhancing stealthiness without compromising effectiveness.
    
    \item We evaluate Sigil in U-SFL across two datasets and four model architectures. The results demonstrate the framework's fidelity, robustness, and stealthiness against the evaluated gradient anomaly detection method. The implementation is publicly available at \url{https://anonymous.4open.science/r/sfl_awm-EEFC/}.
\end{itemize}

The remainder of this article is organized as follows. Section~II reviews related work. Section~III formulates the threat model and design goals. Section~IV presents the proposed Sigil framework. Section~V evaluates its performance. Finally, Section~VI concludes the article and discusses future research directions.

\section{Related Work}
This section reviews existing watermarking methods for the DNN and FL settings, and shows why traditional approaches become inapplicable in the U-SFL setting.

\subsection{DNN Watermarking}
Model watermarking primarily includes backdoor watermarking \cite{217591, 272262, 10646835, DBLP:conf/ndss/00020YHQ025, 10.1145/3196494.3196550}, parameter watermarking \cite{10.1145/3078971.3078974, DBLP:conf/icml/LiuWZ21, 10.1145/3442381.3450000, 10038500, 287178, 298288} and activation watermarking \cite{10.1145/3297858.3304051}. Among them, backdoor watermarking embeds identity information into a model's input--output behavior using a secret trigger set, while parameter watermarking encodes a secret bit string directly into the model's parameter space. Both approaches assume control over the training data, labels, or model parameters. These assumptions do not hold in our U-SFL setting, in which the server cannot access client data or labels and has no control over client-side model parameters.

Compared with the above two categories, activation watermarking is conceptually closer to our problem scenario. This approach embeds watermark information into the activations of intermediate network layers, rather than into data or model parameters. A representative example is DeepSigns \cite{10.1145/3297858.3304051}, which introduces a label-aware mechanism that embeds watermark information into preset class-center vectors, encouraging activations of samples from the same class to cluster around their corresponding centers. However, DeepSigns' dependence on label information makes it incompatible with the U-SFL setting, where the server cannot access such information. Furthermore, as highlighted in a recent systematization-of-knowledge (SoK) study \cite{9833693}, DeepSigns faces limitations in fidelity, robustness, and scalability to complex tasks, making it insufficient for modern IP protection needs.

\subsection{Watermarking in FL and SFL}
FL is the most extensively studied DCML paradigm. Existing FL watermarking research provides useful insights for distributed IP protection. FL watermarking schemes can be divided into client-side embedding and server-side embedding approaches.

Client-side embedding schemes \cite{9847383, 9658998, 10.1145/3630636} assume clients are trusted or semi-honest and perform watermark embedding locally. This assumption contradicts our malicious client threat model. A malicious client may refuse to embed a watermark and even attempt to remove or obfuscate it using full knowledge of the scheme.

Server-side embedding schemes, such as Waffle and FedTracker\cite{9603498, 10504977}, leverage the FL server's authority to directly access, modify, and embed watermarks into the global model before distributing it in each training round. However, this design presupposes server access to the global model parameters---an assumption fundamentally incompatible with U-SFL, where the server has no visibility into client-side model components.

Most closely related to our work, RISE \cite{11593831} addresses ownership verification for both clients and the server in SFL. It combines a server-side feature-based watermark embedded in server-held model parameters with client-side backdoor watermarks introduced through private trigger samples. Sigil addresses a different protection requirement: enabling the server to embed its own ownership credential into a client-held model segment without direct access to its parameters or reliance on the client to execute watermark embedding. We study this requirement in U-SFL, where labels and loss computation also remain on the clients. The distinction therefore lies in the protected model segment, the party initiating the ownership claim, and the capabilities available for watermark embedding.

These studies provide foundations for ownership verification in distributed learning. However, embedding a server-owned credential into client-side model segments without direct parameter access or client-side watermarking cooperation remains a distinct challenge in U-SFL.

\section{Problem Formulation}

This section first formally defines the threat model and  specifies the capabilities of each party. Then it outlines the core design objectives of Sigil.

\subsection{Threat Model}
Based on the motivation  articulated in the Introduction, we define our threat model within the context of U-SFL, as detailed below.

\noindent\textit{Defender Assumptions}

\noindent\textbf{Role.} The defender is the server in the U-SFL framework.

\noindent\textbf{Goal.} The server is the legitimate owner of the global model. Its objective is to enforce a watermark as an IP claim in the client-side model.

\noindent\textbf{Capabilities and Knowledge.} The server possesses its server-side model and has access to the full activation and gradient streams. However, it \textit{cannot} access the client's local data, labels, or model parameters.

\noindent\textit{Adversary Assumptions}

\noindent\textbf{Role.} The adversary is a malicious client in the U-SFL framework.

\noindent\textbf{Goal.} The adversary aims to engage in IP infringement by obtaining a watermark-free version of the client-side model. Their goal is to detect and remove the embedded watermark while maintaining the model's utility, thereby enabling unauthorized distribution or claiming false ownership.

\noindent\textbf{Knowledge.} The adversary inherently possesses white-box access to their local model parameters, training data, and labels. They are aware of the Sigil watermarking algorithm but do not have the secret watermark parameters of the server $K=(M, b)$, where $M$ denotes the embedding matrix and $b$ denotes the watermark string.

\noindent\textbf{Capabilities.} To achieve its goal, the adversary can exert the following methods.
\begin{itemize}
    \item \textbf{Training-Time Gradient Detection}: The adversary can deploy gradient anomaly detection methods (e.g., SplitOut \cite{erdougan2024splitout}) to inspect the server's returned gradients.
    \item \textbf{Training-Time Noise Injection}: The adversary may attempt to disrupt the watermark embedding process by adding noise to the gradients during training.
    \item \textbf{Post-hoc Removal Attacks}: The adversary can apply various known watermark removal attacks---such as fine-tuning, pruning, and quantization, attempting to remove the watermark without degrading main task performance.
\end{itemize}

\subsection{Design Goals}
Sigil must satisfy fidelity, stealthiness, and robustness as its core design objectives. Furthermore, it must support reliable embedding and verification, impose low computational and storage overhead, and provide sufficient capacity to uniquely identify ownership.
\begin{itemize}
    \item \textbf{Fidelity}: Watermark embedding must have a negligible impact on the main task's performance.
    \item \textbf{Stealthiness}: The watermarking mechanism must be statistically indistinguishable from a benign training process so that it can evade  anomaly detection.
    \item \textbf{Robustness}: The watermark must resist various removal attacks defined in the threat model, maintaining its effectiveness.
\end{itemize}

\section{Method}

This section describes the proposed Sigil framework, including design rationale, implementation details, and a geometric interpretation.

\subsection{Design Rationale}

We begin from the U-SFL workflow and the role-capability constraints. Although the server cannot access the client's local data, labels, or model parameters, it observes the smashed activations uploaded by the client and controls the gradients returned to the client. Since the client's model update relies entirely on this gradient, it constitutes the only feasible channel for the server to influence the client's learning.

Inspired by how SL attack techniques like feature-space hijacking attacks \cite{10.1145/3460120.3485259} exploit the backward gradient channel for malicious purposes, we repurpose this mechanism for defense. Specifically,
we utilize the server-visible activation space as the embedding carrier and formulate the watermark as a predefined statistical constraint within this space, rendering it independent of both data and labels. During training, the server computes a watermark loss in parallel with the main-task loss, combines the resulting watermark gradient with the task gradient, and returns the merged gradient to the client. This gradient-injection procedure embeds the watermark without altering the normal training workflow.

A malicious client can deploy gradient detection methods to detect anomalies caused by the watermark in the gradient. To counter the threats of gradient anomaly detection, we adaptively clip the norm of the watermark gradient relative to the task gradient. This adaptive clipping makes the statistical features of the watermark gradient difficult to distinguish from the normal gradient flow, enhancing the scheme's stealthiness. As a result, even if a malicious client knows the scheme's design, it cannot detect the watermark gradient without impairing the optimization of the main task, thus enforcing mandatory embedding.

In summary, our core approach is: \textbf{using gradient injection as the embedding mechanism and the statistical properties of the activation layer as the embedding target, thereby realizing implicit and mandatory watermarking.}

\begin{table}[t]
  \centering
  \caption{Summary of notation.}
  \label{tab:notation}
  \begin{tabularx}{\columnwidth}{@{}lY@{}}
    \toprule
    \textbf{Symbol} & \textbf{Description} \\
    \midrule
    
    $t$ & Global communication round. \\
    $i$ & Client index. \\
    
    $f_{c,i}^{(t)}, \theta_{c,i}^{(t)}$ & Client-side model and its parameters at round $t$. \\
    $f_{s,i}^{(t)}, \theta_{s,i}^{(t)}$ & Server-side model and its parameters at round $t$. \\
    $f'_c, \theta'_c$ & Suspicious client-side model for verification. \\
    
    $X_i, y_i$ & A batch of input data and private labels for client $i$. \\
    $A_i^{(t)}, A_{\text{flat}, i}^{(t)}$ & Smashed data (activation) from client $i$ and its flattened form, respectively. \\
    $S_i^{(t)}$ & Server's intermediate output sent back to client $i$. \\
    $G_{\text{initial}, i}^{(t)}$ & Initial gradient ($\partial \mathcal{L}_{\text{main}} / \partial S_i^{(t)}$) sent from client $i$. \\
    
    $M, b$ & Watermark projection matrix and target bits. \\
    $k$ & The bit-length of the watermark. \\
    $\mathcal{L}_{\text{main}}$ & Main task loss function (e.g., Cross-Entropy). \\
    $\mathcal{L}_{\text{wm}}$ & Watermark loss function (BCE). \\
    $G_{\text{main}, i}^{(t)}$ & Gradient of $\mathcal{L}_{\text{main}}$ w.r.t. client activation $A_i^{(t)}$. \\
    $G_{\text{wm}, i}^{(t)}$ & Gradient of $\mathcal{L}_{\text{wm}}$ w.r.t. client activation $A_i^{(t)}$. \\
    $G_{\text{wm}, i}^{\prime (t)}$ & Watermark gradient after adaptive clipping. \\
    $G_{\text{final}, i}^{(t)}$ & The final (composite) gradient sent to client $i$. \\

    $\lambda$ & Watermark strength (scaling ratio) for adaptive gradient clipping. \\
    $E$ & Number of local epochs for client training. \\
    
    \bottomrule
  \end{tabularx}
\end{table}

\subsection{The Sigil Watermarking Framework}

This section presents the implementation of the Sigil framework, which comprises three core phases: watermark initialization, embedding, and verification. The notation used throughout these processes is summarized in Table \ref{tab:notation}.

\begin{figure*}[t] 
  \centering
  \includegraphics[width=0.8\textwidth]{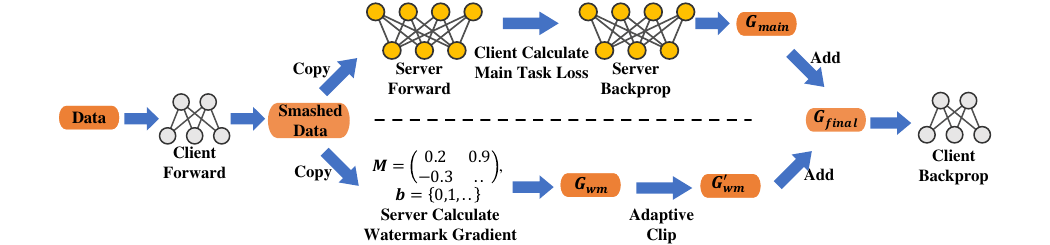}
  \caption{Sigil watermark embedding during U-SFL training.}
  \label{fig:system}
\end{figure*}

\subsubsection{Watermark Initialization} Before the training process begins, the server generates the secret parameter pair $K = (M, b)$ as its ownership credential, consisting of an embedding matrix $M$ and a watermark string $b$.

The embedding matrix $M$ is an $\mathbb{R}^{d \times k}$ matrix with elements sampled independent and identically distributed (IID) from a standard normal distribution $\mathcal{N}(0, 1)$. As a Gaussian matrix, $M$ is full-rank with high probability, which ensures the randomness and stability of the watermark subspace. It serves to project the high-dimensional activation $A_{\text{flat}}$ into a fixed, $k$-dimensional watermark space. 

The watermark string $b$ is a secret bit string of length $k$, sampled uniformly at random from $\{0, 1\}^k$. This binary representation can directly encode information and is well-suited for the subsequent Binary Cross-Entropy (BCE) loss function.

\subsubsection{Watermark Embedding via Gradient Injection} The watermark embedding process is seamlessly integrated into the standard U-SFL workflow. All watermark-related operations are performed solely on the server side and remain entirely invisible to the client.

Sigil leverages these secret parameters ($M, b$) to define the watermark as a statistical constraint on the server-visible activation space. Specifically, through gradient injection, Sigil forces the client's model to generate activations ($A_{flat}$) that statistically match the target bit string $b$ when projected by $M$. The injected gradient is therefore formulated to minimize this projection difference.

A key challenge in this process is maintaining stealthiness. Simply adding the raw watermark gradient $G_{\text{wm}}$ to the main task gradient $G_{\text{main}}$ presents a stealthiness problem \cite{DBLP:conf/ndss/Fu0ZHZJ00023, erdougan2024splitout}. In the early stages of training, before the watermark is well embedded, the watermark loss is often much larger than the main task loss, causing the norm of $G_{\text{wm}}$ to be abnormally large. This would make the statistical features (like the norm) of the final gradient deviate significantly from the normal distribution, creating a detectable anomaly signal for a malicious client.

To address the challenge, we employ an adaptive norm clipping strategy that directly regulates the gradient magnitude instead of reweighting the losses. This method dynamically links the upper bound of the watermark gradient's norm to the main task gradient's norm, controlled by a hyperparameter $\lambda$, yielding a clipped gradient $G_{\text{wm}}^{\prime (i, t)}$.

Finally, the server produces a composite gradient $G_{\text{final}}^{(i, t)}$ and sends it back to client $i$. From the client's perspective, $G_{\text{final}}^{(i, t)}$ is statistically indistinguishable from a standard gradient. Without access to the clean reference gradient, the client cannot detect the subtle watermark perturbations. Therefore, the client is compelled to apply them, thereby being forced to embed the watermark.

Algorithm \ref{alg:embedding} and Fig.~\ref{fig:system} illustrate how Sigil performs watermark embedding within a complete U-SFL forward and backpropagation pass. For clarity, these illustrations focus on the core embedding logic of a single pass, omitting the details of multiple training rounds and the final model aggregation on both sides.

The process begins when the client performs its local forward pass and sends its activation $A_i^{(t)}$ to the server (Line 2). Upon receiving this activation, the server proceeds along two parallel branches: the main task computation (upper branch in Fig.~\ref{fig:system}) and the watermark task computation (lower branch).

For the main task, the server follows the standard U-SFL protocol: it completes the forward pass, returns the logits $S_i^{(t)}$ to the client (Line 4), and subsequently receives the initial gradient $G_{\text{initial},i}^{(t)}$ (Line 6). The server then uses this gradient to perform backpropagation, updating its server-side model and calculating the main task's gradient with respect to the client's activation,  $G_{\text{main}}^{(i, t)}$ (Lines 7-8).

In parallel, for the watermark task, the server uses its secret parameters $K=(M, b)$ to compute the watermark loss $\mathcal{L}_{\text{wm}}$. This loss is defined as the BCE between the target watermark string $b$ and the predicted bit string:

\begin{equation}
\mathcal{L}_{wm} = \text{BCE}(\text{sigmoid}(A_{flat}^{(i, t)} \cdot M), b)\text{.}
\label{eq:lwm}
\end{equation}

Based on this loss, the server calculates the corresponding raw watermark gradient $G_{\text{wm}}^{(i, t)}$ (Line 9). The server then applies the adaptive norm clipping mechanism (Line 10) to generate the stealthy, clipped gradient $G_{\text{wm}}^{\prime (i, t)}$:

\begin{equation}
G_{\text{wm}}^{\prime (i, t)} = \min\left(1, \lambda \cdot \frac{\|G_{main}^{(i, t)}\|_2}{\|G_{wm}^{(i, t)}\|_2 + \epsilon}\right) \cdot G_{wm}^{(i, t)}\text{,}
\label{eq:gwm}
\end{equation}

where $\epsilon$ is a small constant to avoid division by zero. Finally, the server combines the gradients from both the main task and the watermark task (Line 11) to obtain the composite gradient $G_{\text{final}}^{(i, t)}$:

\begin{equation}
G_{\text{final}}^{(i, t)} = G_{\text{main}}^{(i, t)} + G_{\text{wm}}^{\prime (i, t)}\text{.}
\label{eq:gfinal}
\end{equation}

This final gradient is then sent to the client (Line 12), which uses it to perform local backpropagation and update the client-side model.

\begin{algorithm}[!t]
\caption{U-SFL Training with Sigil Embedding (Single Batch Step within Round $t$)}
\label{alg:embedding}
\begin{algorithmic}[1]
    \Statex \textbf{Input:} Client $i$'s data batch $(X_i, y_i)$ and current model $\theta_{c,i}^{(t)}$; 
    \Statex \quad Server's current model $\theta_{s,i}^{(t)}$, watermark parameters $(M, b)$ and $\lambda$.
    \Statex \textbf{Output:} Locally updated client model $\theta_{c,i}^{(t)}$ and server model $\theta_{s,i}^{(t)}$.

    \Statex \quad \textbf{Client $i$ Executes:}
    \State $A_i^{(t)} \leftarrow f_{c,i}^{(t)}(X_i)$
    \State Send $A_i^{(t)}$ to Server

    \Statex \quad \textbf{Server Executes:}
    \State $S_i^{(t)} \leftarrow f_{s,i}^{(t)}(A_i^{(t)})$
    \State Send $S_i^{(t)}$ back to Client $i$
    
    \Statex \quad \textbf{Client $i$ Executes:}
    \State $G_{\text{initial}, i}^{(t)} \leftarrow \frac{\partial \mathcal{L}_{\text{main}}(S_i^{(t)}, y_i)}{\partial S_i^{(t)}}$
    \State Send $G_{\text{initial}, i}^{(t)}$ to Server 

    \Statex \quad \textbf{Server Executes:}
    \State $\theta_{s,i}^{(t+1)} \leftarrow \text{Update}(\theta_{s,i}^{(t)}, G_{\text{initial}, i}^{(t)})$  
    \State $G_{\text{main}, i}^{(t)} \leftarrow \text{Backprop}(G_{\text{initial}, i}^{(t)}, A_i^{(t)})$
    \State $G_{\text{wm}, i}^{(t)} \leftarrow \frac{\partial \mathcal{L}_{\text{wm}}(A_i^{(t)}, M, b)}{\partial A_i^{(t)}}$
    \State $G_{\text{wm}, i}^{\prime (t)} \leftarrow \min(1, \lambda \cdot \frac{\|G_{\text{main}, i}^{(t)}\|_2}{\|G_{\text{wm}, i}^{(t)}\|_2 + \epsilon}) \cdot G_{\text{wm}, i}^{(t)}$ 
    \State $G_{\text{final}, i}^{(t)} \leftarrow G_{\text{main}, i}^{(t)} + G_{\text{wm}, i}^{\prime (t)}$
    \State Send final gradient $G_{\text{final}, i}^{(t)}$ to Client $i$

    \Statex \quad \textbf{Client $i$ Executes:}
    \State $\theta_{c,i}^{(t+1)} \leftarrow \text{Update}(\theta_{c,i}^{(t)}, G_{\text{final}, i}^{(t)})$ 
    
\end{algorithmic}
\end{algorithm}

\subsubsection{Watermark Verification} The watermark verification process allows the owner to confirm ownership of a potentially leaked model. A key advantage of this verification method is its data-agnostic nature: the verifier does not need access to any private original training data.

The verification process assumes white-box access to the target model $f'_c$, specifically the ability to perform a forward pass and obtain the activation output $A'$ at the split point. Using these activations and the secret watermark parameters $K = (M, b)$, the verifier checks whether the embedded statistical constraint is preserved.

We quantify the verification result using the Watermark Success Rate (WSR). Ownership is confirmed if the computed WSR exceeds a preset threshold $\tau$. This statistically justified threshold is determined empirically, as described in Section~\ref{sec:metrics}.

The specific procedure outlined in Algorithm \ref{alg:verification} is as follows:

The verifier first generates a set of $N$ random input tensors $X_{\text{rand}}$, (e.g., from $\mathcal{N}(0, 1)$) (Line 1). It then performs a forward pass on the suspect model to get the activations $A'$ (Line 2), flattens them (Line 3), and projects them using the secret matrix $M$ to get the predicted probabilities $\hat{b}$ (Line 4). These probabilities are subsequently rounded to extract the predicted bit string $B'$ (Line 5).

Finally, the $WSR_{\text{verify}}$ is calculated (Line 6) by comparing the predicted bit string $B'$ against the target watermark $b$, aggregated across all samples and bits:

\begin{equation}
WSR = \frac{\sum_{i=1}^{N} \sum_{j=1}^{k} \mathbb{I}(b'_{ij} == b_j)}{N \times k}\text{.}
\label{eq:wsr}
\end{equation}

Here, $N$ is the number of random samples used for verification, $k$ is the watermark length (defined in Table \ref{tab:notation}), $b'_{ij}$ is the predicted $j$-th bit for the $i$-th sample, and $b_j$ is the $j$-th bit of the target watermark. $\mathbb{I}(\cdot)$ is the indicator function. This calculated WSR is then compared to the threshold $\tau$ to return the final boolean verification result (Lines 7-11).

\begin{algorithm}[!t]
\caption{Sigil Post-hoc Data-Free Verification}
\label{alg:verification}
\begin{algorithmic}[1]
    \Statex \textbf{Verifier Executes:}
    \Statex \textbf{Input:} Suspect model parameters $\theta'_c$; Watermark parameters $(M, b)$; Threshold $\tau$ 
    \Statex \textbf{Output:} A boolean value indicating whether passing the ownership verification process. 

    \State Generate a set of $N$ random tensors $X_{\text{rand}}$ (e.g., from $\mathcal{N}(0, 1)$)

    \State $A' \leftarrow f_c(X_{\text{rand}}, \theta'_c)$ 
    \State $A'_{\text{flat}} \leftarrow \text{Flatten}(A')$
    \State $\hat{b} \leftarrow \text{sigmoid}(A'_{\text{flat}} \cdot M)$

    \State $B' \leftarrow \text{round}(\hat{b})$
    \State $WSR_{\text{verify}} \leftarrow \frac{\sum_{i=1}^{N} \sum_{j=1}^{k} \mathbb{I}(B'_{ij} == b_j)}{N \times k}$ 
    \If{$WSR_{\text{verify}} > \tau$}
        \State \textbf{return} True
    \Else
        \State \textbf{return} False
    \EndIf
\end{algorithmic}
\end{algorithm}

\subsection{Geometric Interpretation of the Embedding Mechanism}

This section provides a geometric interpretation of the Sigil framework's embedding process that explains its effectiveness and high fidelity, aligning with two key experimental observations. We model the gradient updates as occurring in two distinct subspaces. The first is the watermark subspace ($S_{\text{wm}}$),  a fixed $k$-dimensional subspace spanned by the column vectors of the secret matrix $M$. The second is the main task subspace ($S_{\text{main}}$), which is determined by the data distribution and the main-task loss.

The random initialization of $M$ motivates a geometric interpretation of the interaction between the watermark and main-task gradients. In high-dimensional spaces, independently sampled random directions tend to exhibit weak alignment. Specifically, for two independently and uniformly sampled unit vectors in $\mathbb{R}^d$, the expected squared cosine similarity satisfies \cite{VershyninHDP2}:

\begin{equation}
E[\cos^2(\theta)] = \text{Var}(\cos(\theta)) = \frac{1}{d}\text{.}
\label{eq:cos}
\end{equation}

This geometric perspective suggests that small watermark updates with weak alignment to the main-task gradient may introduce limited interference with task optimization. Consistent with this intuition, the experiments in Section~\ref{sec:fidelity_exp} and Fig.~\ref{fig:cos} report mean cosine similarities close to zero between $G_{\text{main}}$ and $G_{\text{wm}}$, alongside limited changes in main-task accuracy.

This perspective also helps interpret the effectiveness of Sigil at small watermark strengths. Repeated watermark updates encourage the projected activations to match the same target bit string, allowing a weak watermark signal to accumulate over training. The main-task updates, in contrast, are driven by the prediction objective and need not consistently reinforce or oppose this target. This interpretation is consistent with the results in Section~\ref{sec:fidelity_exp}, where effective embedding ($WSR > 99\%$) is achieved at $\lambda=0.01$, with the watermark gradient norm restricted to at most 1\% of the main-task gradient norm.

\section{Evaluation}

This section provides a comprehensive evaluation of the proposed Sigil watermarking framework in terms of fidelity, stealthiness, robustness, and key design choices, including the split strategy and watermark length.

\subsection{Experiment Setup}

\subsubsection{Datasets, Models, and Settings} \label{sec:setup} We evaluate Sigil on two benchmark datasets: CIFAR-10 (using ResNet-18 and VGG11) and Tiny-ImageNet (using MobileNetV2 and DenseNet-121). We simulate a 10-client IID U-SFL setting. Unless otherwise specified, our default configuration uses ResNet-18 on CIFAR-10 with the split point placed after the third residual block. The watermark strength is set to $\lambda = 0.1$ and the watermark length to $k = 50$.

\subsubsection{Implementation and Training Details}
All experiments were conducted on a server running Ubuntu 22.04. We utilized a single NVIDIA RTX 5090 GPU for all training and evaluation. The implementation was built using Python 3.12, PyTorch 2.8.0, and CUDA 13.0. 

For the training protocol, all models were trained for a total of 80 global communication rounds. In each round, every client performed $E=2$ local epochs of training on its private data. We used a global batch size of 256.

For optimization, we employed the stochastic gradient descent (SGD) optimizer with a momentum of 0.9 and weight decay of 5e-4. The learning rate schedule utilized a combination of a 5-round linear warmup followed by cosine annealing. The base learning rate was set to $1 \times 10^{-1}$, and the learning rate annealed down to a minimum of $1 \times 10^{-4}$ by the end of training.

\subsubsection{Evaluation Metrics} \label{sec:metrics} We use Top-1 Accuracy (Acc) to measure main-task performance and WSR (defined in Eq.~\ref{eq:wsr}) to assess watermark effectiveness. To establish a statistically significant verification threshold $\tau$, we adapt the methodology of Lukas \textit{et al}.~\cite{9833693}. Specifically, we empirically evaluate our ResNet-18 baseline by testing 100 randomly generated watermark parameters against 30 independently trained clean models, and observe that the resulting WSR values follow a normal distribution. From this distribution, we determine that a 5-sigma significance level ($p\text{-value} < 3 \times 10^{-7}$) corresponds to a WSR of $68\%$. Accordingly, we conservatively set the verification threshold to $\tau = 70\%$, exceeding the 5-sigma level and providing strong statistical confidence for all subsequent claims.

\subsection{Fidelity}
\label{sec:fidelity_exp}

This section evaluates the fidelity of Sigil by comparing a clean model (trained without watermarking) with watermarked models (WM) under different watermark strengths ($\lambda$). 

\begin{figure}[t]
    \centering
    \includegraphics[width=\columnwidth]{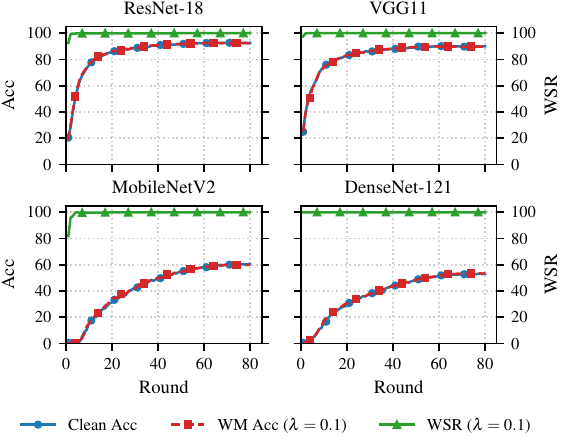}
    \caption{Fidelity and Effectiveness Evaluation on All Tasks.}
    \label{fig:effandfid}
\end{figure}

Fig.~\ref{fig:effandfid} illustrates the training dynamics of Sigil. Across all four task settings, the accuracy of the watermarked models nearly coincides with that of the clean baseline, indicating high fidelity.

Table~\ref{tab:fidelity_effectiveness} presents detailed results supporting this observation. The accuracy difference between the watermarked models and the clean baseline remains within a narrow range of $\pm 0.6\%$ for all tasks. Furthermore, accuracy does not exhibit a noticeable downward trend as $\lambda$ increases, suggesting that watermark strength has no significant negative impact on main-task performance.

\begin{table*}[t]
\centering
\footnotesize
\caption{Quantitative results for fidelity, effectiveness, and gradient orthogonality.Each cell shows Top: Main Task Acc (\%) / Middle: WSR (\%) / Bottom: Mean Cosine Similarity.
}
\label{tab:fidelity_effectiveness}

\begin{tabular}{@{}l c ccc@{}} 
\toprule
\multirow{2}{*}{\textbf{Model / Dataset}} & \multirow{2}{*}{\textbf{Baseline (Clean)}} & \multicolumn{3}{c}{\textbf{Sigil Watermark ($\lambda$ value)}} \\
\cmidrule(l){3-5} 
& & $\lambda=0.01$ & $\lambda=0.1$ & $\lambda=1.0$ \\
\midrule

\textbf{ResNet-18 / CIFAR-10} &
\makecell{92.6\% \\ - \\ -} &
\makecell{92.4\%(-0.2) \\ 99.1\% \\ 2.5e-6} &
\makecell{92.4\%(-0.2) \\ 100\% \\ -2.1e-6} &
\makecell{92.8\%(+0.2) \\ 100\% \\ 1.6e-6} \\
\addlinespace 

\textbf{VGG11 / CIFAR-10} &
\makecell{89.9\% \\ - \\ -} &
\makecell{89.9\% \\ 99.9\% \\ -1.1e-4} &
\makecell{89.7\%(-0.2) \\ 100\% \\ -2.0e-5} &
\makecell{89.9\% \\ 100\% \\ -1.9e-5} \\
\addlinespace

\textbf{MobileNetV2 / Tiny-ImageNet} &
\makecell{60.4\% \\ - \\ -} &
\makecell{60.5\%(+0.1) \\ 99.2\% \\ -1.8e-5} &
\makecell{60.0\%(-0.4) \\ 100\% \\ 5.5e-6} &
\makecell{60.1\%(-0.3) \\ 100\% \\ 7.3e-6} \\
\addlinespace

\textbf{DenseNet-121 / Tiny-ImageNet} &
\makecell{52.9\% \\ - \\ -} &
\makecell{53.2\%(+0.3) \\ 100\% \\ -2.4e-6} &
\makecell{53.5\%(+0.6) \\ 100\% \\ 3.6e-6} &
\makecell{52.9\% \\ 100\% \\ -7.7e-6} \\

\bottomrule
\end{tabular}
\end{table*}

Notably, the watermark embedding process is highly effective. Even at an extremely low strength of $\lambda = 0.01$, the WSR consistently reaches $99\%$ or higher, far exceeding the verification threshold of $\tau = 70\%$. This demonstrates that Sigil remains effective even under low gradient strength, which aligns with the geometric analysis.

\begin{figure}[t]
    \centering
    \includegraphics[width=0.8\columnwidth]{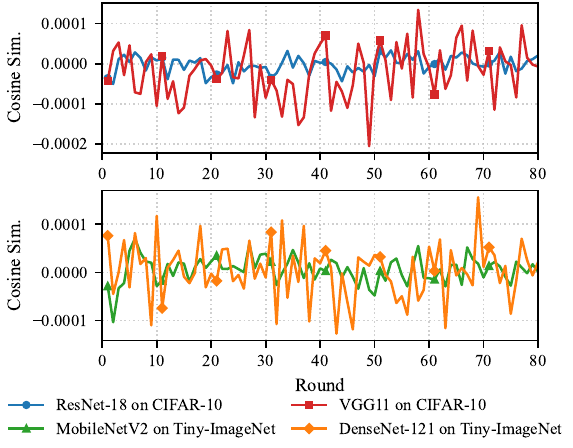}
    \caption{Gradient Orthogonality Analysis (Cosine Similarity).}
    \label{fig:cos}
\end{figure}

The gradient alignment results provide an empirical perspective on this fidelity. As shown in Table~\ref{tab:fidelity_effectiveness}, the mean cosine similarity between $G_{\text{wm}}$ and $G_{\text{main}}$ remains close to zero across the evaluated configurations. Fig.~\ref{fig:cos} shows that the roundwise values fluctuate around zero on the order of $10^{-4}$. These observations are consistent with the limited changes in main-task accuracy.

\begin{figure*}[t] 
    \centering
    \includegraphics[width=0.8\textwidth]{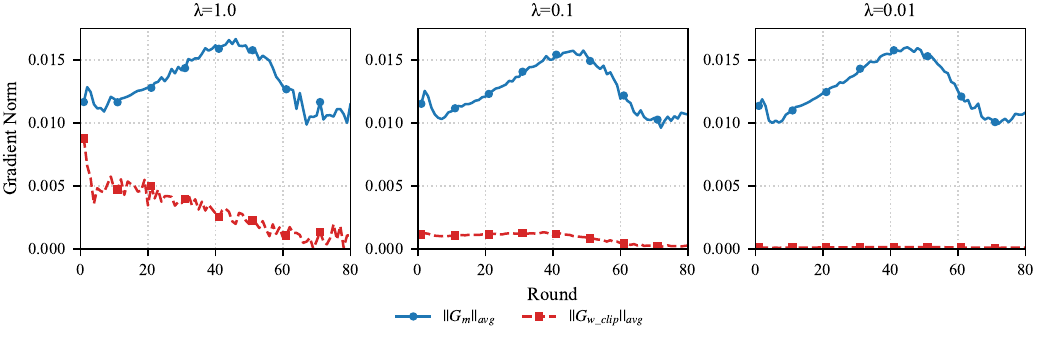}
    \caption{Gradient norm dynamics during training (ResNet-18/CIFAR-10). Adaptive clipping at $\lambda=0.1$ and $0.01$ keeps the watermark gradient norm well below the main-task gradient norm.}
    \label{fig:grad_comp}
\end{figure*}

Fig.~\ref{fig:grad_comp} further illustrates the gradient norm dynamics. At $\lambda=1.0$, the watermark gradient has a larger magnitude, whereas smaller strengths ($\lambda=0.1$ and $0.01$) keep its norm well below that of the main-task gradient. Together with the observed gradient alignment and task accuracy, these results support the effectiveness of adaptive clipping in limiting the watermark's impact on the main task.

\subsection{Stealthiness} \label{sec:stealthiness}
This section first provides a brief overview of gradient anomaly detection methods and then evaluates the stealthiness of Sigil by comparing the gradient streams of the clean baseline and the watermarked models under the detection scheme.

In SL and its variants, the activation space and the gradient channel constitute the primary interfaces between clients and the server and therefore become focal points for both attacks and defenses.  Prior work has shown that adversaries can exploit the server's control over the gradient channel for malicious purposes \cite{10.1145/3460120.3485259, DBLP:conf/aaai/YuZZPW24, 10188623}, and corresponding defenses have been developed to detect anomalies in gradient streams. Since Sigil also operates through and perturbs the gradient channel, such detectors also pose a direct challenge to its stealthiness.

Early methods such as SplitGuard \cite{10.1145/3559613.3563198} relied on active injection, whereas Pinocchio's Nose \cite{DBLP:conf/ndss/Fu0ZHZJ00023} adopted a passive detection strategy. More recently, SplitOut \cite{erdougan2024splitout} improved upon heuristic approaches by constructing a robust reference distribution from normal gradients.

SplitOut constructs a reference distribution from clean gradients and identifies anomalous sample gradients. In our implementation, gradients from client 0 are checked in non-overlapping windows of 256 samples, and an alert is raised when more than 128 gradients in a window are anomalous. For visualization, we report the total number of anomalous gradients collected over each training round; the ablation tables report the mean of these roundwise totals.

\begin{figure}[t]
    \centering
    \includegraphics[width=0.8\columnwidth]{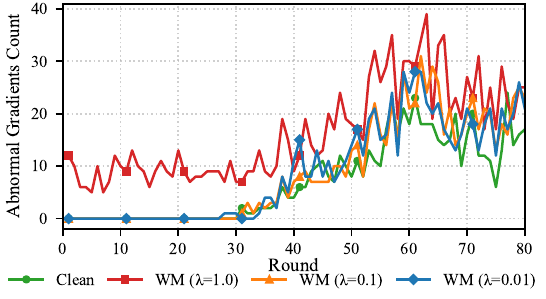}
    \caption{Stealthiness Evaluation against SplitOut}
    \label{fig:splitout}
\end{figure}

As shown in Fig.~\ref{fig:splitout}, the clean baseline exhibits fluctuations in roundwise anomaly counts. The outlier count for WM ($\lambda = 1.0$) is consistently higher than this clean baseline. In contrast, the outlier curves for WM ($\lambda = 0.1$) and WM ($\lambda = 0.01$) remain tightly aligned with the clean baseline's fluctuation range. Even the peak outlier count for WM ($\lambda = 1.0$) (around 40) remains far below the typical batch-level detection threshold (128 for our batch size of 256).

The results confirmed the high stealthiness of the Sigil framework. The experiments clearly demonstrate that watermark strength $\lambda$ is the dominant factor: when $\lambda$ is large (e.g., 1.0), the watermark introduces detectable statistical deviations, whereas smaller values (e.g., 0.1) allow Sigil to effectively suppress its gradient impact and evade detection. Combined with the findings in Section \ref{sec:fidelity_exp}, Sigil's ability to remain highly effective even at extremely low strengths (e.g., $\lambda = 0.01$) indicates that the practical feasible range for $\lambda$ is wide, underscoring the scheme's utility.

\subsection{Robustness}
This section evaluates the robustness of Sigil against training-time noise injection and post-hoc watermark removal attacks. 

\subsubsection*{Robustness Against Training-Time Noise Injection} 
\begin{figure}[t]
    \centering
    \includegraphics[width=0.8\columnwidth]{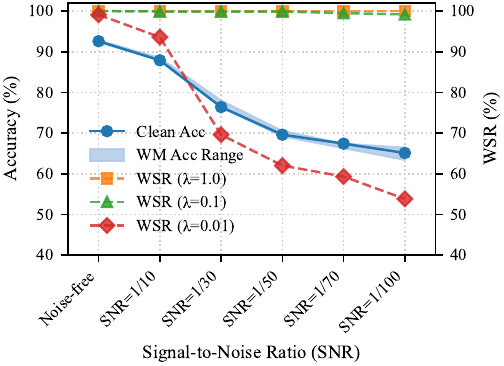}
    \caption{Robustness against Training-Time Noise Injection (SNR).}
    \label{fig:nsr}
\end{figure}

To simulate a client interfering with the $G_{\text{final}}$ transmission, we adopt ideas from prior works \cite{mao2023secure, 2022arXiv220104018G} to inject Gaussian noise, quantified using the Signal-to-Noise Ratio (SNR). As shown in Fig.~\ref{fig:nsr}, we plot results across decreasing SNR values (i.e., increasing noise) from "Noise-free" to $1/100$, where lower SNR corresponds to stronger noise.

As shown in Fig.~\ref{fig:nsr}, the accuracy of both the clean model and the watermarked models steadily decreases as noise strength increases (i.e., as SNR decreases). Accuracy drops from approximately 90\% to 70\% as SNR decreases from noise-free to $1/50$, and further declines to around 65\% at $1/100$.

In contrast, the Sigil watermark exhibits strong robustness to gradient noise. The WSR for $\lambda = 1.0$ remains at 100\% even under the highest noise level (SNR = $1/100$), while the WSR for $\lambda = 0.1$ remains at 99.2\% under the same conditions. The $\lambda = 0.01$ watermark is more affected, dropping to 69.6\% (below the verification threshold $\tau = 70\%$) at SNR = $1/30$. At this noise level, the main-task accuracy also decreases substantially, illustrating the trade-off between watermark suppression and model utility.

\subsubsection*{Robustness Against Post-hoc Removal Attacks} We evaluate Sigil against fine-tuning, pruning, and quantization, comparing it to simulated backdoor (BDWM) and parameter (PWM) watermark baselines \cite{9847383, DBLP:conf/ndss/00020YHQ025}.

\begin{table*}[h]
\centering
\footnotesize

\caption{Robustness against post-training removal attacks (ResNet-18/CIFAR-10). Each cell shows Top: Acc (\%) / Bottom: WSR (\%).
}
\label{tab:robustness_removal}
\begin{tabular}{@{}l l c cc ccc@{}}
\toprule

\multirow{2}{*}{\textbf{Attack Type}} & \multirow{2}{*}{\textbf{Strength}} & \multirow{2}{*}{\textbf{Clean}} & \multicolumn{2}{c}{\textbf{Baselines (Sim.)}} & \multicolumn{3}{c}{\textbf{Sigil ($\lambda$ value)}} \\
\cmidrule(l){4-5} \cmidrule(l){6-8}

& & \textbf{(Acc)} & \textbf{BDWM} & \textbf{PWM} & $\lambda=1.0$ & $\lambda=0.1$ & $\lambda=0.01$ \\
\midrule

\textbf{No Attack} & \textbf{(None)}
& 92.6\%
& \makecell{92.1\% \\ 99.9\%}
& \makecell{92.4\% \\ 100\%}
& \makecell{92.8\% \\ 100\%}
& \makecell{92.4\% \\ 100\%}
& \makecell{92.4\% \\ 99.1\%} \\
\midrule 

\multirow{2}{*}{\textbf{Fine-tuning}} & 50 steps
& 91.5\%
& \makecell{91.4\% \\ 98.1\%(-1.8)}
& \makecell{91.6\% \\ 100\%}
& \makecell{91.8\% \\ 100\%}
& \makecell{91.8\% \\ 100\%}
& \makecell{91.4\% \\ 99.0\%(-0.1)} \\
 
& 100 steps
& 91.3\%
& \makecell{91.3\% \\ 95.6\%(-4.3)}
& \makecell{91.5\% \\ 100\%}
& \makecell{91.6\% \\ 100\%}
& \makecell{91.5\% \\ 100\%}
& \makecell{91.2\% \\ 99.0\%(-0.1)} \\
\midrule 

\multirow{2}{*}{\textbf{Pruning}} & Ratio 60\%
& 52.0\%
& \makecell{65.9\% \\ 75.8\%(-24.1)}
& \makecell{60.3\% \\ 100\%}
& \makecell{48.2\% \\ 100\%}
& \makecell{53.7\% \\ 99.7\%(-0.3)}
& \makecell{53.6\% \\ 98.3\%(-1.6)} \\
 
& Ratio 80\%
& 14.8\%
& \makecell{10.5\% \\ 0\%(-99.9)}
& \makecell{11.7\% \\ 100\%}
& \makecell{11.7\% \\ 97.9\%(-2.1)}
& \makecell{10.5\% \\ 94.6\%(-5.4)}
& \makecell{10.0\% \\ 87.5\%(-12.4)} \\
\midrule 

\multirow{3}{*}{\textbf{Quantization}} & FP16
& 92.6\%
& \makecell{92.2\% \\ 99.9\%}
& \makecell{92.4\% \\ 100\%}
& \makecell{92.8\% \\ 100\%}
& \makecell{92.4\% \\ 100\%}
& \makecell{92.4\% \\ 99.1\%} \\
\addlinespace

& INT8
& 92.6\%
& \makecell{92.2\% \\ 99.9\%}
& \makecell{92.4\% \\ 100\%}
& \makecell{92.8\% \\ 100\%}
& \makecell{92.4\% \\ 100\%}
& \makecell{92.4\% \\ 99.1\%} \\
\addlinespace

& INT4
& 89.6\%
& \makecell{90.4\% \\ 99.8\%(-0.1)}
& \makecell{90.2\% \\ 98.0\%(-2)}
& \makecell{91.1\% \\ 100\%}
& \makecell{91.4\% \\ 100\%}
& \makecell{90.5\% \\ 99.0\%(-0.1)} \\

\bottomrule
\end{tabular}
\end{table*}

The results are presented in Table~\ref{tab:robustness_removal}. PWM exhibits extremely high robustness (WSR $\ge 98.0\%$) across all attacks due to its inherent mechanism. In contrast, BDWM is more fragile, with its WSR decreasing to 95.6\% under fine-tuning and failing completely under 80\% pruning.

Sigil demonstrates excellent robustness to both fine-tuning and quantization. Under 100 steps of strong fine-tuning and extreme INT4 quantization, the WSR for all Sigil configurations remains virtually unaffected (WSR $\ge 99.0\%$ for $\lambda = 0.01$), matching the performance of the most robust PWM baseline. For pruning, Sigil also exhibits strong watermark retention: at an 80\% pruning rate, the WSR for $\lambda = 1.0$ drops only slightly to 97.9\%, while $\lambda = 0.1$ drops to 94.6\%. The weakest $\lambda = 0.01$ model experiences the largest reduction to 87.5\%, yet this remains well above the verification threshold of 70\%. At this pruning rate, main-task accuracy falls close to chance level, indicating that the watermark persists even under severe model degradation. Overall, these results demonstrate strong watermark retention under the evaluated attacks, with the fine-tuning and quantization settings also preserving main-task accuracy.

\subsection{Ablation Studies}

This section conducts ablation studies on the split point, the length of watermark,  client count and Non-IID data distribution.

\subsubsection{Ablation on Split Strategy} As shown in Table~\ref{tab:ablation_split}, without attacks, Acc remains within 92.1\%--92.4\% across all split points, while WSR reaches 99.8\% or higher. This indicates that Sigil can be flexibly deployed at different depths of the model without impacting its baseline fidelity and effectiveness.

Stealthiness improves significantly at deeper split points. The SplitOut outlier count, while fluctuating in shallow layers (16.7 for Layer 1 and 20.44 for Layer 2), drops substantially to 8.95 (Layer 3) and 0 (Layer 4). This trend is broadly consistent for robustness. Under extreme noise interference ($SNR=1/100$), the WSR surges from 73.7\% (Layer 1) to 99.5\% (Layer 4). Similarly, under 80\% pruning, the WSR at Layers 2-4 ($\ge 94.6\%$) is markedly superior to Layer 1 (88.6\%).

In summary, the results show that embedding at deeper locations significantly improves stealthiness and robustness. However, deeper split points also impose a higher computational overhead on the client. Our choice of Layer 3 as the baseline is a comprehensive trade-off between robustness, stealthiness, and computational cost.

\begin{table*}[ht]
\centering
\footnotesize
\setlength{\tabcolsep}{6pt}

\caption{Ablation study on split point strategy (ResNet-18/CIFAR-10, $\lambda=0.1$). All cells (except SplitOut) show Top: Acc (\%) / Bottom: WSR (\%). FT, PR, and QT denote fine-tuning, pruning, and quantization, respectively. The SplitOut column reports the mean roundwise total of anomalous gradients for client 0 (lower is better).}
\label{tab:ablation_split}

\begin{tabular}{@{}l c c c ccc@{}} 
\toprule

\multirow{3}{*}{\textbf{Split Point}} & \textbf{Fidelity} & \textbf{Stealthiness} & \multicolumn{4}{c}{\textbf{Robustness}} \\
\cmidrule(l){2-2} \cmidrule(l){3-3} \cmidrule(l){4-7} 

& \multirow{2}{*}{\textbf{No Attack}} & \multirow{2}{*}{\makecell{\textbf{SplitOut} ($\downarrow$)}} & \textbf{Training-time} & \multicolumn{3}{c}{\textbf{Post-hoc}} \\
\cmidrule(l){4-4} \cmidrule(l){5-7} 

& & & \makecell{Noise\\(snr=$1/100)$} & \makecell{FT\\(100 steps)} & \makecell{PR\\(80\%)} & \makecell{QT\\(INT4)} \\
\midrule

\textbf{Layer 1}
& \makecell{92.4\% \\ 99.8\%} & 16.7
& \makecell{87.1\% \\ 73.7\%(-26.1)} & \makecell{91.3\% \\ 99.8\%} & \makecell{14.3\% \\ 88.6\%(-11.2)} & \makecell{90.1\% \\ 99.8\%} \\
\addlinespace

\textbf{Layer 2}
& \makecell{92.2\% \\ 99.9\%} & 20.44
& \makecell{79.9\% \\ 96.3\%(-3.6)} & \makecell{91.0\% \\ 99.9\%} & \makecell{10.0\% \\ 97.5\%(-2.4)} & \makecell{89.8\% \\ 99.9\%} \\
\addlinespace

\textbf{Layer 3}
& \makecell{92.4\% \\ 100\%} & 8.95
& \makecell{64.6\% \\ 99.2\%(-0.8)} & \makecell{91.5\% \\ 100\%} & \makecell{10.5\% \\ 94.6\%(-5.4)} & \makecell{91.4\% \\ 100\%} \\
\addlinespace

\textbf{Layer 4}
& \makecell{92.1\% \\ 100\%} & 0
& \makecell{39.5\% \\ 99.5\%(-0.5)} & \makecell{91.6\% \\ 99.9\%(-0.1)} & \makecell{14.6\% \\ 94.8\%(-5.2)} & \makecell{90.7\% \\ 100\%} \\
\bottomrule
\end{tabular}
\end{table*}

\subsubsection{Ablation on Watermark Length} We conduct this ablation at the default split point (Layer 3), where the activation tensor has a total dimension of 16,384, providing a theoretical reference for capacity.

As shown in Table \ref{tab:ablation_length}, Sigil demonstrates significant watermark capacity. In the No Attack column, the WSR remains at 100\% even as the watermark length $k$ increases to 3,000. However, this high capacity comes at the cost of other key attributes. As $k$ increases, the Acc slightly drops from 92.4\% to 91.7\%. Robustness against standard attacks also declines: under 80\% pruning, WSR falls from 94.6\% ($k=50$) to the 84.8\%-88.5\% range ($k \ge 500$). 

The most severe cost of increasing $k$ is to stealthiness. As shown in the table's SplitOut column, the average outlier count surges from 8.95 ($k=50$) to 592.39 ($k=3000$). This makes high-capacity Sigil statistically trivial to detect, failing the stealthiness requirement.

Watermark length $k$ is a critical design choice. While larger watermarks retain high WSR in the absence of attacks, the evaluated settings show increased anomaly counts, reduced WSR after pruning, and a modest decline in task accuracy. These results highlight the need to select $k$ by jointly considering fidelity, stealthiness, and robustness.

\begin{table*}[t]
\centering
\footnotesize
\setlength{\tabcolsep}{8pt}

\caption{Ablation study on watermark length (k) (ResNet-18/CIFAR-10, $\lambda=0.1$). All cells (except SplitOut) show Top: Acc (\%) / Bottom: WSR (\%). FT, PR, and QT denote fine-tuning, pruning, and quantization, respectively. The SplitOut column reports the mean roundwise total of anomalous gradients for client 0 (lower is better). Alert status is determined separately by the window-level criterion described in Section~\ref{sec:stealthiness}.}
\label{tab:ablation_length}

\begin{tabular}{@{}l c c ccc@{}} 
\toprule
\multirow{3}{*}{\textbf{Length ($k$)}} & \textbf{Fidelity} & \textbf{Stealthiness} & \multicolumn{3}{c}{\textbf{Robustness}} \\
\cmidrule(l){2-2} \cmidrule(l){3-3} \cmidrule(l){4-6} 
& \multirow{2}{*}{\textbf{No Attack}} & \multirow{2}{*}{\makecell{\textbf{SplitOut} ($\downarrow$)}} & \multicolumn{3}{c}{\textbf{Post-hoc}} \\
\cmidrule(l){4-6} 
& & & \makecell{FT\\(100 steps)} & \makecell{PR\\(80\%)} & \makecell{QT\\(INT4)} \\
\midrule

$k$=50
& \makecell{92.4\% \\ 100\%} & 8.95
& \makecell{91.5\% \\ 100\%} & \makecell{10.5\% \\ 94.6\%(-5.4)} & \makecell{91.4\% \\ 100\%} \\
\addlinespace

$k$=500
& \makecell{92.2\% \\ 100\%} & 37.41
& \makecell{90.9\% \\ 100\%} & \makecell{10.0\% \\ 86.6\%(-13.4)} & \makecell{89.2\% \\ 100\%} \\
\addlinespace

$k$=1000
& \makecell{92.2\% \\ 100\%} & 65.41
& \makecell{91.0\% \\ 100\%} & \makecell{11.4\% \\ 84.8\%(-15.2)} & \makecell{89.7\% \\ 100\%} \\
\addlinespace

$k$=2000
& \makecell{91.5\% \\ 100\%} & \makecell{165.4\\(Not Alerted)}
& \makecell{90.5\% \\ 99.9\%(-0.1)} & \makecell{10.0\% \\ 85.9\%(-14.1)} & \makecell{89.5\% \\ 99.9\%(-0.1)} \\
\addlinespace

$k$=3000
& \makecell{91.7\% \\ 100\%} & \makecell{592.39\\(Alerted)}
& \makecell{90.7\% \\ 99.8\%(-0.2)} & \makecell{10.0\% \\ 88.5\%(-11.5)} & \makecell{87.0\% \\ 99.9\%(-0.1)} \\

\bottomrule
\end{tabular}
\end{table*}

\subsubsection{Impact of Client Counts}

Table \ref{tab:adaptive_clients} shows that for watermarking strengths $\lambda=1$ and $\lambda=0.1$, the WSR remains at 100\%. At the weakest strength ($\lambda=0.01$), the WSR exhibits a slight decline as $c$ increases, reaching 89.0\% at $c=100$. We hypothesize this decline is due to insufficient embedding, a consequence of the minimal data available per client at this low strength. Nevertheless, this WSR remains well above the 70\% success threshold, even as Acc declines significantly (owing to the reduced data volume per client).

\begin{table}[t]
\centering
\footnotesize
\caption{Sigil effectiveness under different client counts(c) and watermark strengths $\lambda$. Each cell shows Top: Acc (\%) / Bottom: WSR (\%).}
\label{tab:adaptive_clients}
\begin{tabular}{@{}l ccc@{}}
\toprule
\multirow{2}{*}{\textbf{Client Counts}} & \multicolumn{3}{c}{\textbf{Watermark Strength $\lambda$}} \\
\cmidrule(l){2-4} 
& $\lambda=1$ & $\lambda=0.1$ & $\lambda=0.01$ \\
\midrule

\textbf{$c=10$} &
\makecell{92.8 \\ 100.0} &
\makecell{92.4 \\ 100.0} &
\makecell{92.4 \\ 99.1(-0.9)} \\
\addlinespace

\textbf{$c=20$} &
\makecell{89.8 \\ 100.0} &
\makecell{88.9 \\ 100.0} &
\makecell{89.8 \\ 98.4(-1.6)} \\
\addlinespace 

\textbf{$c=50$} &
\makecell{83.9 \\ 100.0} &
\makecell{84.4 \\ 100.0} &
\makecell{84.2 \\ 97.3(-2.7)} \\
\addlinespace

\textbf{$c=100$} &
\makecell{74.1 \\ 100.0} &
\makecell{73.8 \\ 100.0} &
\makecell{72.1 \\ 89.0(-11.0)} \\

\bottomrule
\end{tabular}
\end{table}

\begin{table}[t]
\centering
\footnotesize
\caption{Watermark embedding results under different Non-IID levels $\beta$ and watermark strengths $\lambda$. Each cell shows Top: Acc (\%) / Bottom: WSR (\%).}
\label{tab:adaptive_noniid}
\begin{tabular}{@{}l ccc@{}}
\toprule
\multirow{2}{*}{\textbf{Non-IID $\beta$}} & \multicolumn{3}{c}{\textbf{Watermark Strength $\lambda$}} \\
\cmidrule(l){2-4} 
& $\lambda=1$ & $\lambda=0.1$ & $\lambda=0.01$ \\
\midrule

\textbf{IID} &
\makecell{92.8 \\ 100.0} &
\makecell{92.4 \\ 100.0} &
\makecell{92.4 \\ 99.1(-0.9)} \\
\addlinespace

\textbf{$\beta=1.0$} &
\makecell{90.8 \\ 100.0} &
\makecell{91.3 \\ 100.0} &
\makecell{90.6 \\ 99.2(-0.8)} \\
\addlinespace 

\textbf{$\beta=0.5$} &
\makecell{90.1 \\ 100.0} &
\makecell{90.5 \\ 100.0} &
\makecell{88.9 \\ 99.4(-0.6)} \\
\addlinespace

\textbf{$\beta=0.1$} &
\makecell{66.3 \\ 100.0} &
\makecell{77.2 \\ 100.0} &
\makecell{70.9 \\ 99.5(-0.5)} \\

\bottomrule
\end{tabular}
\end{table}

\begin{table}[t]
\centering
\footnotesize
\caption{Watermark embedding results under different unbalanced levels $\sigma$ and watermark strengths $\lambda$. Each cell shows Top: Acc (\%) / Bottom: WSR (\%).}
\label{tab:adaptive_unbalanced}
\begin{tabular}{@{}l ccc@{}}
\toprule
\multirow{2}{*}{\textbf{Unbalanced $\sigma$}} & \multicolumn{3}{c}{\textbf{Watermark Strength $\lambda$}} \\
\cmidrule(l){2-4} 
& $\lambda=1$ & $\lambda=0.1$ & $\lambda=0.01$ \\
\midrule

\textbf{Balanced} & 
\makecell{92.8 \\ 100.0} &
\makecell{92.4 \\ 100.0} &
\makecell{92.4 \\ 99.1(-0.9)} \\
\addlinespace

\textbf{$\sigma=0.5$} &
\makecell{92.5 \\ 100.0} &
\makecell{92.4 \\ 100.0} &
\makecell{92.5 \\ 98.9(-1.1)} \\
\addlinespace 

\textbf{$\sigma=1.0$} &
\makecell{92.4 \\ 100.0} &
\makecell{92.3 \\ 100.0} &
\makecell{92.2 \\ 99.2(-0.8)} \\
\addlinespace

\textbf{$\sigma=2.0$} &
\makecell{92.2 \\ 100.0} &
\makecell{92.6 \\ 100.0} &
\makecell{92.2 \\ 98.6(-1.4)} \\

\bottomrule
\end{tabular}
\end{table}

\subsubsection{Impact of Data Heterogeneity} Real-world FL scenarios often involve heterogeneous data. We evaluated Sigil's robustness against two types of heterogeneity, including data Distribution Heterogeneity (Non-IID) and Data Quantity Heterogeneity (Unbalanced).

\begin{itemize}
    \item \textbf{Data Distribution Heterogeneity (Non-IID):} We simulate this using a Dirichlet distribution ($Dir(\beta)$). The parameter $\beta$ controls the degree of Non-IID; a smaller $\beta$ indicates greater disparity among clients. We test $\beta \in \{1.0, 0.5, 0.1\}$, where $\beta=1.0$ represents a moderate skew and $\beta=0.1$ represents extreme Non-IID (i.e., each client's data consists almost exclusively of one or two classes).
    \item \textbf{Data Quantity Heterogeneity (Unbalanced):} We simulate this by controlling the standard deviation $\sigma$ of client sample sizes, which follows a log-normal distribution. A larger $\sigma$ signifies a greater imbalance in data quantity across clients. We test $\sigma \in \{0.5, 1.0, 2.0\}$, where $\sigma=0.5$ represents a slight fluctuation and $\sigma=2.0$ indicates an extreme disparity in data volume (e.g., 10\% of clients holding 60-80\% of the data).
\end{itemize}

Tables \ref{tab:adaptive_noniid} and \ref{tab:adaptive_unbalanced} demonstrate that Sigil is highly robust to data heterogeneity. For all tested $\beta$ (down to 0.1) and $\sigma$ (up to 2.0) values, the WSR remains above 98\%. This indicates that Sigil can reliably embed and verify watermarks despite skewed data class distributions or severe data quantity imbalances among clients.

In summary, this analysis confirms that Sigil remains effective under realistic U-SFL scenarios, including a large number of clients and severe data heterogeneity (in distribution and quantity), which demonstrates its practical feasibility.

\section{Conclusion and Future Work}
In this paper, we addressed the challenge of enforcing IP protection for capability-limited servers in U-SFL environments. We formalized the threat model and identified the limitations of traditional watermarking methods in protecting client-side model segments under restricted server capabilities. To address this challenge, we proposed Sigil, a server-enforced watermarking framework that leverages gradient injection to embed statistical constraints into the client-side model. Additionally, through an adaptive gradient clipping mechanism, Sigil balances robustness, fidelity, and stealthiness against anomaly detection. Extensive experiments in U-SFL validate the effectiveness of our approach under the evaluated settings, supporting its potential application to model ownership verification in edge IoT services.

Despite its promising performance, several directions merit future exploration. First, like most activation-based and parameter-based schemes, Sigil is currently vulnerable to feature permutation attacks \cite{9833693}. Permuting the neurons at the split point can disrupt the statistical alignment of specific activation channels without affecting main task performance. Developing permutation-invariant embedding targets remains a challenging open problem for the community. Furthermore, Sigil is specifically designed to verify the ownership of model parameters against unauthorized leakage, rather than preventing model extraction attacks that aim to mimic model functionality. Future work could explore integrating Sigil with complementary defenses to provide more comprehensive protection against both parameter leakage and functionality mimicry.

\bibliographystyle{IEEEtran}
\bibliography{references} 

\begin{IEEEbiography}[{\includegraphics[width=1in,height=1.25in,clip,keepaspectratio]{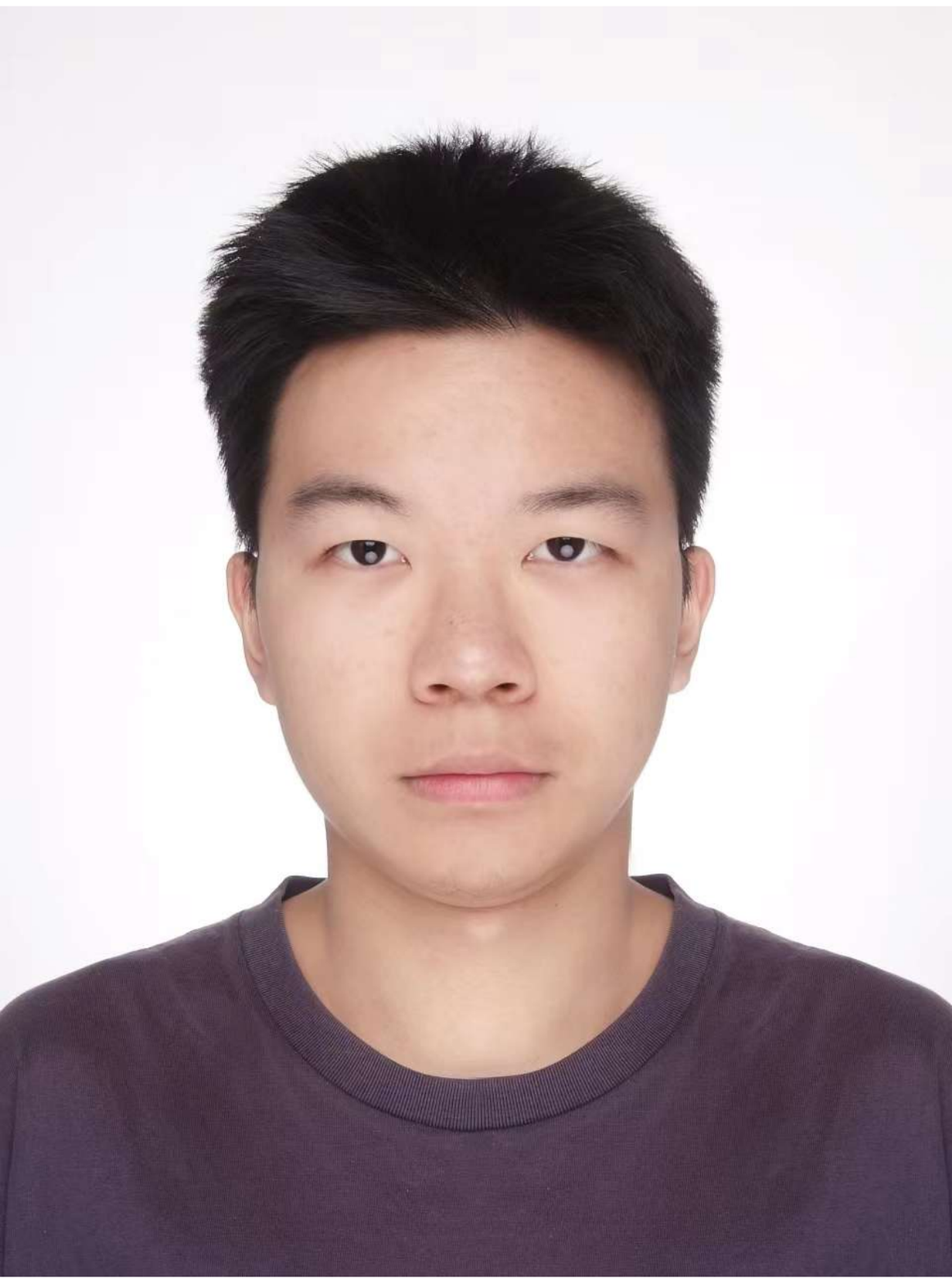}}]{Zhengchunmin Dai}
received the B.Eng. degree in software engineering from East China Normal University, China, in 2024. 

He is currently pursuing the master's degree in software engineering with East China Normal University, China. His research interests include AI security and model intellectual property protection.
\end{IEEEbiography}

\begin{IEEEbiography}[{\includegraphics[width=1in,height=1.25in,clip,keepaspectratio]{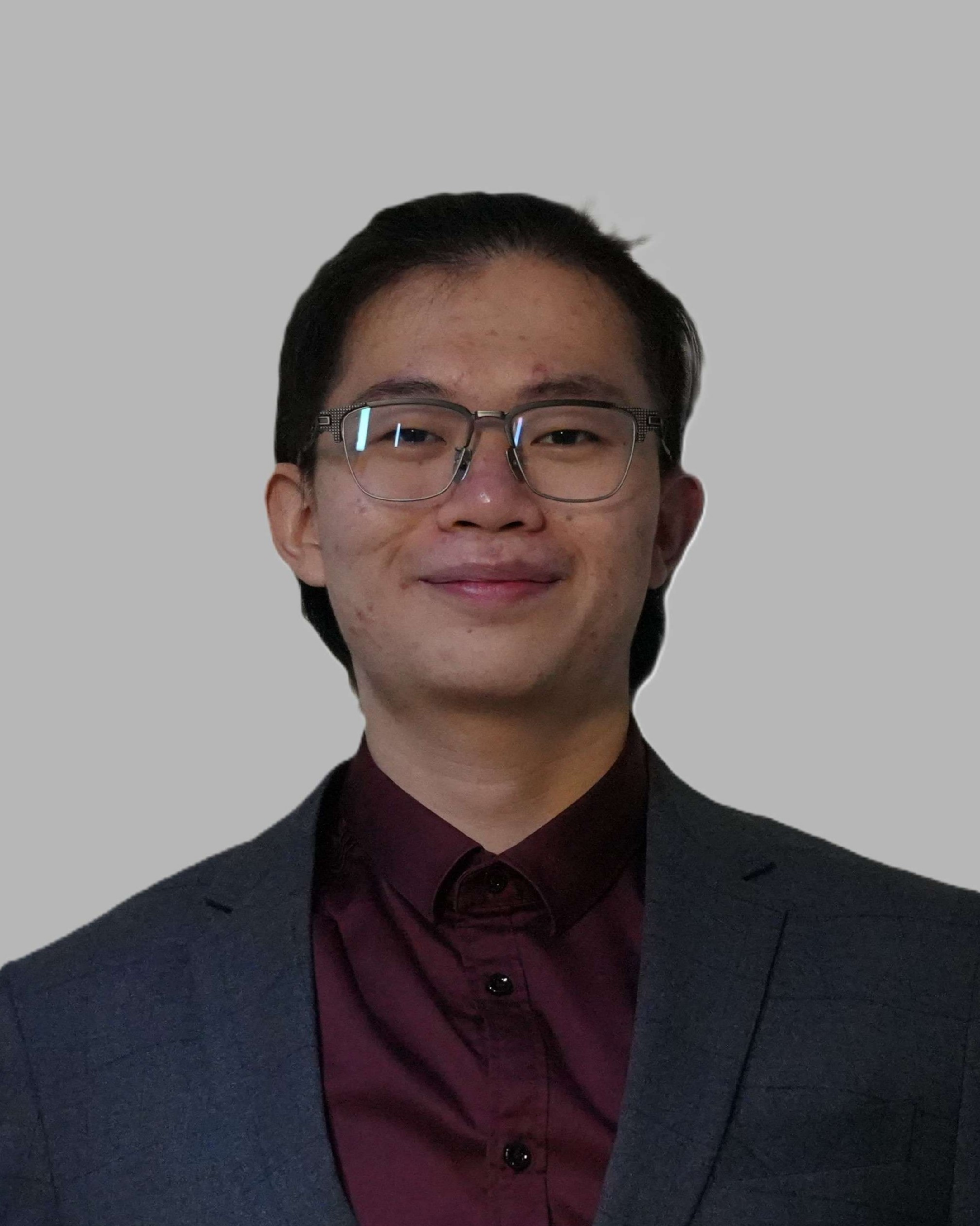}}]{Jiaxiong Tang}
received the B.Eng. degree in network engineering from Dalian University of Technology, China, in 2024. 

He is currently pursuing the master's degree in software engineering with East China Normal University, China. His research interests include trustworthy AI, privacy-preserving machine learning, and model intellectual property protection.
\end{IEEEbiography}

\begin{IEEEbiography}[{\includegraphics[width=1in,height=1.25in,clip,keepaspectratio]{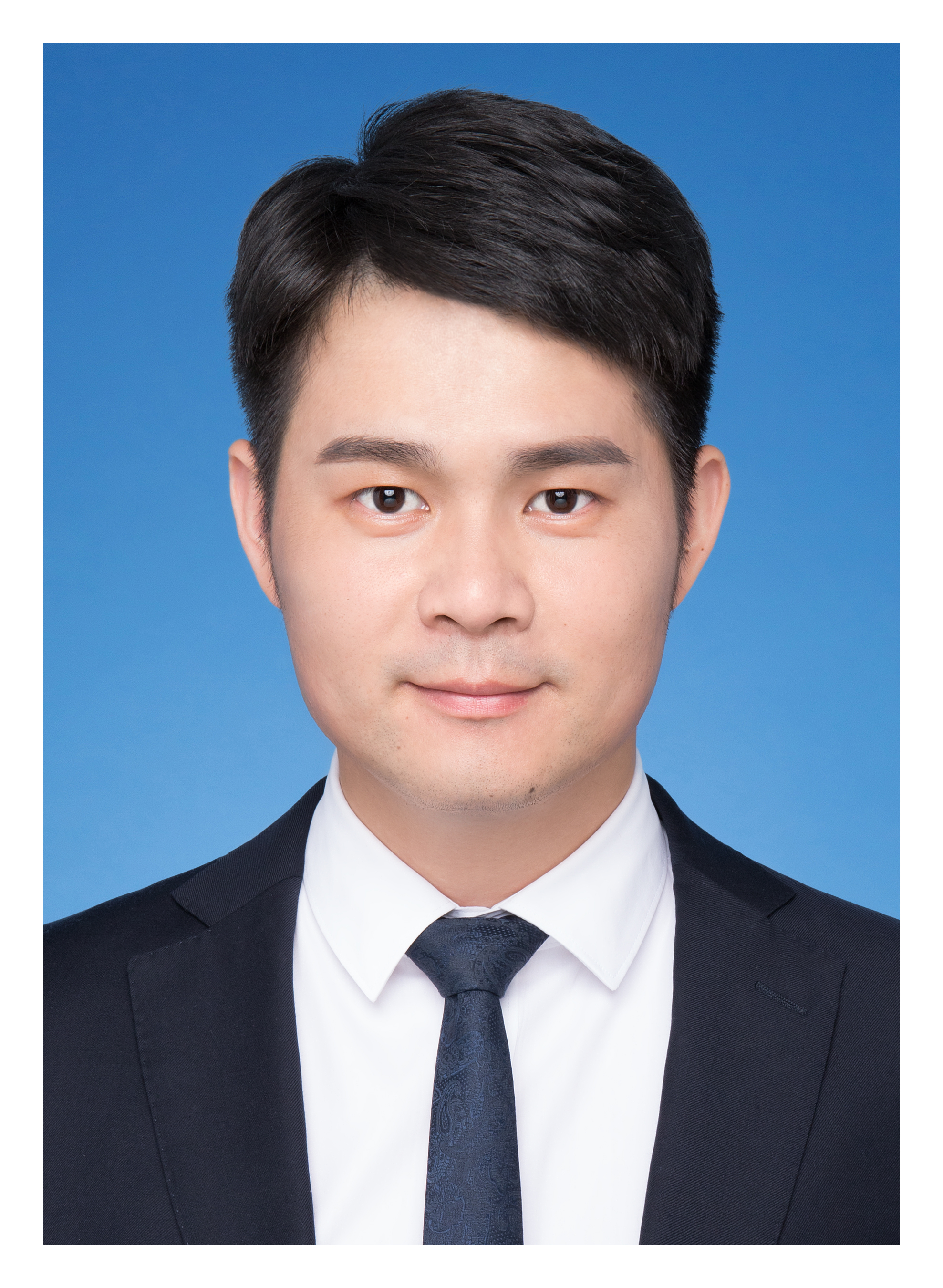}}]{Peng Sun} received the B.E. degree in automation from Tianjin University, China, in 2015, and his Ph.D. degree in control science and engineering from Zhejiang University, China, in 2020.

He is currently an Associate Professor with the College of Computer Science and Electronic Engineering, Hunan University, China. His research interests include the Internet of Things, federated learning, security and privacy.
\end{IEEEbiography}

\begin{IEEEbiography}[{\includegraphics[width=1in,height=1.25in,clip,keepaspectratio]{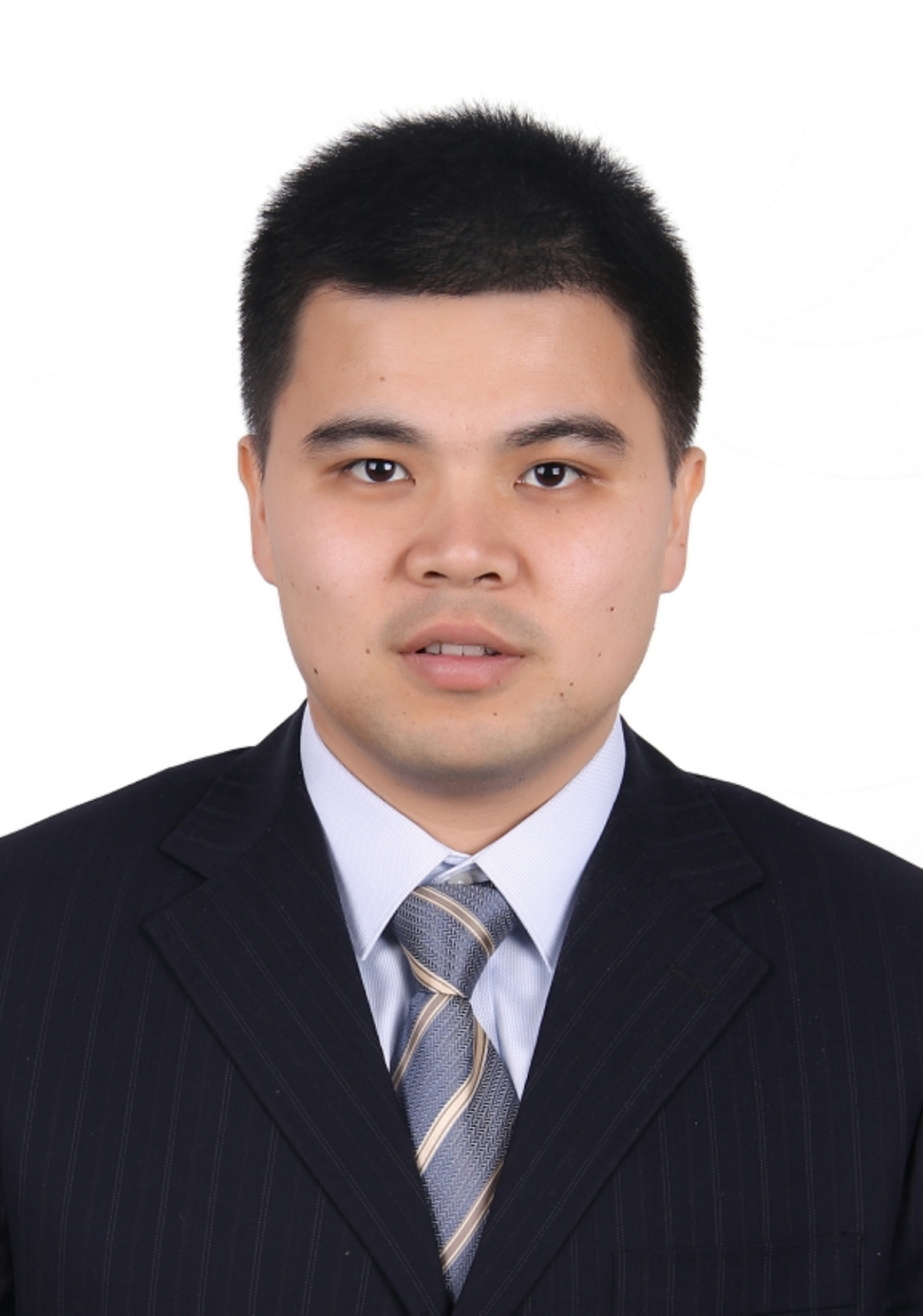}}]{Honglong Chen}
received the M.E. degree in Control Science and Engineering from Zhejiang University, China, in 2008, and the Ph.D degree in Computer Science from The Hong Kong Polytechnic University, Hong Kong, in 2012. He was a Postdoctoral Researcher in the School of CIDSE at Arizona State University from 2015 to 2016. 

He is currently a Professor and Ph.D supervisor with the College of Control Science and Engineering, China University of Petroleum, China. He served as a Guest Editor of IEEE Transactions on Industrial Informatics. His current research interests are in the areas of Internet of Things and AI security. He is a senior member of IEEE and CCF (China Computer Federation) and a member of ACM.
\end{IEEEbiography}

\begin{IEEEbiography}[{\includegraphics[width=1in,height=1.25in,clip,keepaspectratio]{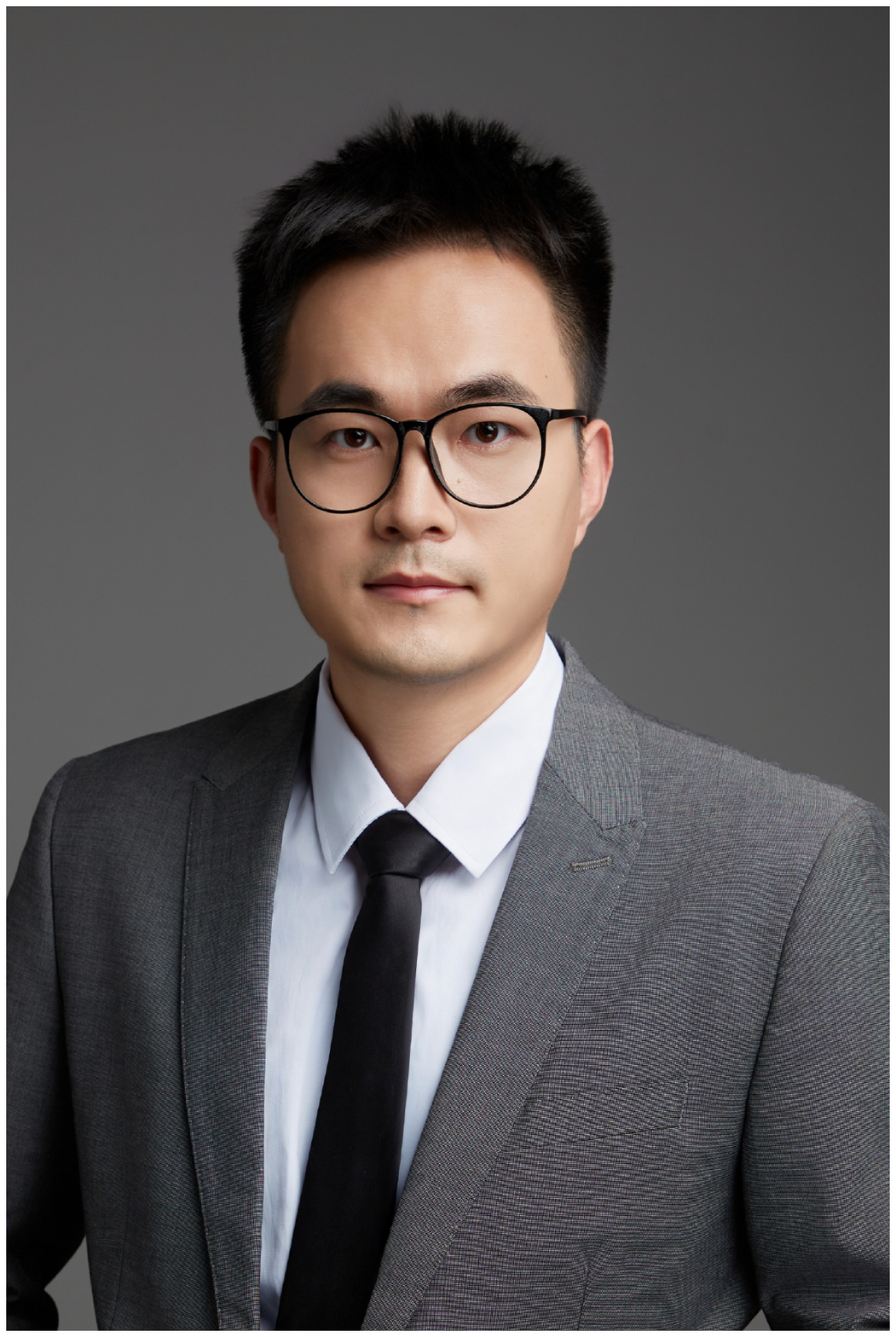}}]{Liantao Wu}
received the B.E. degree in automation from Shandong University, China, in 2012 and his Ph.D. degree in Control Science and Engineering from Zhejiang University, China, in 2017.
 
He is currently an Associate Professor with  the Software Engineering Institute, East China Normal University,
China. His research interests include IoT, edge intelligence and AI security.
\end{IEEEbiography}

\vfill

\end{document}